\documentclass[10pt]{article}

\usepackage{amssymb}
\usepackage{amsmath}

\usepackage[colorlinks=true, linkcolor=blue, citecolor=blue, urlcolor=blue]{hyperref}

\usepackage{graphicx}
\usepackage{booktabs}
\usepackage{multirow}
\usepackage{tabularx}
\usepackage{enumitem}
\usepackage{threeparttable}
\usepackage{balance}
\usepackage{url}

\usepackage{sectsty}
\sectionfont{\fontsize{10}{10}\selectfont}
\subsectionfont{\fontsize{10}{10}\selectfont}

\begin{document}

\title{On Algorithmic Fairness and the EU Regulations}

\author{Jukka Ruohonen \\ University of Southern Denmark \\ \texttt{juk@mmmi.sdu.dk}}

\maketitle

\begin{abstract}
The short paper discusses algorithmic fairness by focusing on non-discrimination
and a few important laws in the European Union (EU). In addition to the EU laws
addressing discrimination explicitly, the discussion is based on the EU's
recently enacted regulation for artificial intelligence (AI) and the older
General Data Protection Regulation (GDPR). Through a theoretical scenario
analysis, on one hand, the paper demonstrates that correcting discriminatory
biases in AI systems can be legally done under the EU regulations. On the other
hand, the scenarios also illustrate some practical scenarios from which legal
non-compliance may follow. With these scenarios and the accompanying discussion,
the paper contributes to the algorithmic fairness research with a few legal
insights, enlarging and strengthening also the growing research domain of
compliance in AI engineering. \\
\\\vspace{20pt}
\end{abstract}

\section{Introduction}

Predictive machine learning models and automated decision-making (ADM) systems
have been widely developed and deployed in recent years. Alongside the
developments and deployments has been a long-standing debate over biases these
models and systems have or may have. Recently, the debate has moved toward
revolving around ``formalists'' and ``theorists'' (for a lack of better terms),
the former seeking to approach biases through formal algorithmic modeling and
the latter connecting fairness to broader legal, philosophical, and ethical
reasoning.\footnote{~Ben Green, `Escaping the Impossibility of Fairness: From
Formal to Substantive Algorithmic Fairness' (2022) 35 \textit{Philosophy \&
  Technology} $<$\url{https://doi.org/10.1007/s13347-022-00584-6}$>$.} The paper
follows the latter group: the goal is to elaborate algorithmic fairness and
practical bias correction in relation to a few important EU regulations.

Before continuing further, three brief terminological remarks should be
made. First, to narrow the scope of concepts and to align with the regulations,
a term ``AI system'' is taken as a synonym with ADM systems based on machine
learning. The scenarios considered in Section~\ref{sec: scenarios} are all
contextualized with AI systems nowadays widely used in recruitment of
employees. Second, the difficult concept of algorithmic fairness is narrowed by
only focusing on discrimination in the recruitment context considered. In other
words, someone might be discriminated by an AI recruitment system because of his
or her religion, gender, mother tongue, or even political opinions. Third, the
other difficult concept of a bias is framed with a conventional statistical bias
involving a non-representative sample with respect to a larger
population.\footnote{~Mitchell, Eric Potash, Solon Barocas, Alexander D'Amour
and Kristian Lum, `Algorithmic Fairness: Choices, Assumptions, and Definitions'
(2021) 8 \textit{Annual Review of Statistics and Its Application} $<$\url{https://doi.org/10.1146/annurev-statistics-042720-125902}$>$.} This
choice simplifies the scenario analysis and aligns it toward the GDPR because the
underlying presumption is that hypothetical developers of an AI recruitment
system would seek to correct biases by collecting further personal data of
prospective employees.

With the partial focus on data protection, the paper contributes to the
discussion on the GDPR's relation to algorithmic fairness. Recently, an argument
was raised that the GDPR might need a new exemption for this particular
case.\footnote{~Marvin van Bekkum and Frederik Zuiderveen Borgesius, `Using
Sensitive Data to Prevent Discrimination by Artificial Intelligence: Does the
GDPR Need a New Exception?' (2023) 48 \textit{Computer Law \& Security Review},
$<$\url{https://doi.org/10.1016/j.clsr.2022.105770}$>$.} As will be elaborated,
the EU's new artificial intelligence regulation has largely addressed this
concern. The theoretical scenario analysis highlights also other scenarios under
which bias correction can be done under the GDPR. From a broader perspective,
the paper also contributes to the growing research domain of compliance in
engineering of AI systems, including with respect to the new and old EU
regulations.\footnote{~Teresa Scantamburlo, Paolo Falcarin, Alberto Veneri,
Alessandro Fabris, Chiara Gallese, Valentina Billa, Francesca Rotolo and
Federico Marcuzzi, 'Software Systems Compliance with the AI Act: Lessons Learned
from an International Challenge' \textit{Proceedings of the 2nd International
  Workshop on Responsible AI Engineering (RAIE)} (2024) Lisbon
$<$\url{https://doi.org/10.1145/3643691.3648589}$>$.} To this end, the opening
Section~\ref{sec: regulations} introduces the EU laws considered in the
theoretical scenario analysis. Then, the analytical approach for the scenario
analysis is briefly discussed in Section~\ref{sec: approach}. After presenting
the scenarios in Section~\ref{sec: scenarios}, a conclusion follows in
Section~\ref{sec: conclusion}.

\section{Regulations}\label{sec: regulations}

There are two EU laws addressing non-discrimination explicitly. In general,
these prohibit discrimination on grounds such as racial or ethnic origin,
religion or belief, age, disability, and sexual orientation.\footnote{~Council
Directive 2000/43/EC of 29 June 2000 Implementing the Principle of Equal
Treatment Between Persons Irrespective of Racial or Ethnic Origin (2000) OJ L
180. Council Directive 2000/78/EC of 27 November 2000 Establishing a General
Framework for Equal Treatment in Employment and Occupation (2000) OJ L 303.} The
legal basis and background come from the EU's fundamental rights charter, which
states that ``\textit{discrimination based on any ground such as sex, race,
  colour, ethnic or social origin, genetic features, language, religion or
  belief, political or any other opinion, membership of a national minority,
  property, birth, disability, age or sexual orientation shall be
  prohibited}''.\footnote{~Article 21 in Charter of Fundamental Rights of the
European Union (2012) C 326/391.} According to a recent evaluation, most member
states have adopted the Charter's mandate and the two laws into their national
jurisdictions, but fragmentation remains a problem and infringement proceedings
are underway against some member states.\footnote{~European Commission, `A
Comparative Analysis of Non-Discrimination Law in Europe 2023: The 27 EU Member
States Compared' (Publications Office of the European Union, Luxembourg 2024)
$<$\url{https://op.europa.eu/en/publication-detail/-/publication/0624900d-e73b-11ee-9ea8-01aa75ed71a1/language-en}$>$.}
Regarding the fragmentation, an important point is that the scope of
discriminating factors vary from a member state to another. Some member states
have narrow listings of factors, while in others listings are extensive,
surpassing the amount of factors coming from the EU laws. For instance, Belgium,
Croatia, Finland, and France have specified also labor union membership as a
factor. This judicial fragmentation may pose a problem for an attempt to
implement algorithmic fairness uniformly across Europe. Another important point
is that already the amount of factors listed in the Charter is enough to remark
a limitation in existing algorithmic fairness research in that it has often
considered fairness only in terms of a dichotomous distinction between two
groups, advantaged and disadvantaged.\footnote{~\textit{Supra note}~2.} In
reality, discrimination may involve multiple factors simultaneously, some of
which may intersect or correlate with each other.

Regarding artificial intelligence, relevant is also the recently enacted
so-called AI Act in the EU.\footnote{~Regulation (EU) 2024/1689 of the European
Parliament and of the Council of 13 June 2024 Laying Down Harmonised Rules on
Artificial Intelligence and Amending Regulations (EC) No 300/2008, (EU) No
167/2013, (EU) No 168/2013, (EU) 2018/858, (EU) 2018/1139 and (EU) 2019/2144 and
Directives 2014/90/EU, (EU) 2016/797 and (EU) 2020/1828 (Artificial Intelligence
Act) (2024) OJ L, 2024/1689.} Although it emphasizes that existing laws are not
affected, including the mentioned non-discrimination laws and data protection
laws, it also states that high-risk AI systems should be rigorously assessed for
potential biases that may lead to discrimination.\footnote{~Recital 45 and
Article~10, paragraphs (2)(f) and (2)(g) in \textit{supra note} 9.}  This
evaluation mandate constitutes the paper's central theme.

Then there is the EU's famous GDPR.\footnote{~Regulation (EU) 2016/679 of the
European Parliament and of the Council of 27 April 2016 on the Protection of
Natural Persons with Regard to the Processing of Personal Data and on the Free
Movement of Such Data, and Repealing Directive 95/46/EC (General Data Protection
Regulation) (2016) OJ L 119.} It contains six potential legal bases for lawful
processing of personal data, among which informed consent is likely the most
used in the private sector.\footnote{~Article 6 and Article 6(a), respectively
in \textit{supra note} 11.}  For the purposes of this paper, relevant is also
the GDPR's notion of special or sensitive categories of personal data. These are
defined to include ``\textit{personal data revealing racial or ethnic origin,
  political opinions, religious or philosophical beliefs, or trade union
  membership, and the processing of genetic data, biometric data for the purpose
  of uniquely identifying a natural person, data concerning health or data
  concerning a natural person's sex life or sexual
  orientation}''.\footnote{~Article 9(1) in \textit{supra note} 11.}  Processing
such data is generally prohibited, although there are ten exceptions to this
prohibition. Even though a reason has not been fully investigated, it may be
that these relaxations partially explain an observation that the prohibition
does not appear high in the empirical rankings of legal reasons for GDPR
enforcement fines.\footnote{~Jukka Ruohonen and Kalle Hjerppe, `The GDPR
Enforcement Fines at Glance' \textit{Information Systems} (2022) 106
$<$\url{https://doi.org/10.1016/j.is.2021.101876}$>$. Marlene Saemann, Daniel
Theis, Tobias Urban and Martin Degeling, 'Investigating GDPR Fines in the Light
of Data Flows', \textit{Proceedings on Privacy Enhancing Technologies}' (2022)
4 $<$\url{https://petsymposium.org/popets/2022/popets-2022-0111.pdf}$>$.} In
any case, the prohibition, even under the exemptions, provides the motivating
kernel for the paper: discrimination based on special categories of personal
data is prohibited and even collection of such data is generally
\text{prohibited---and} yet AI systems should be designed and audited for
potential discrimination involving exactly those data categories. This dilemma
is explicitly present also in the GDPR, which too emphasizes that the special
categories of personal data should not lead to discrimination through biases,
inaccuracies, or other errors.\footnote{~Recital 71 in \textit{supra note} 11.}

\section{Approach}\label{sec: approach}

The approach is based on what might be called argumentative compliance. It has
been a typical approach in software engineering, including its such subdomains
as software architectures and particularly requirements
engineering.\footnote{~Giampaolo Armellin, Annamaria Chiasera, Ivan Jureta,
Alberto Siena and Angelo Susi, `Establishing Information System Compliance: An
Argumentation-Based Framework' \textit{Proceedings of the Fifth International
  Conference on Research Challenges in Information Science (RCIS)} (2011) Gosier
$<$\url{https://doi.org/10.1109/RCIS.2011.6006853}$>$. Kalle Hjerppe, Jukka
Ruohonen and Ville Lepp\"anen, `The General Data Protection Regulation:
Requirements, Architectures, and Constraints' \textit{Proceedings of the 27th
  IEEE International Requirements Engineering Conference (RE)} (2019) Jeju
Island $<$\url{https://doi.org/10.1109/RE.2019.00036}$>$. Boyan Mihaylov, Lucian
Onea and Klaus Marius Hansen, ?Architecture-Based Regulatory Compliance
Argumentation' \textit{Journal of Systems and Software} (2016) 119
$<$\url{https://doi.org/10.1016/j.jss.2016.04.057}$>$.} Drawing on argumentation
theory and formal reasoning, the approach's general background is rich and
flexible.\footnote{~Brigitte Burgemeestre, Joris Hulstijn and Yao-Hua Tan,
  `Value-Based Argumentation for Justifying Compliance' (2011) 19
  \textit{Artificial Intelligence and Law}
  $<$\url{https://doi.org/10.1007/s10506-011-9113-4}$>$.}

However, to simplify the analysis, the focus is only on two common reasoning
patterns: necessary conditions and compensating measures. The former are
generally about assessing whether given measures taken were necessary in a sense
that without them the goals stated would not have been achieved, whereas the
latter is about assessing whether any measure taken could have been replaced by
an alternative measure.\footnote{~\textit{Supra note} 17.} Instead of focusing on
goals, which might be used to assess the GDPR's information security
requirements, necessary conditions are assessed by considering whether a measure
can be reasonably deduced to be legally invalid from which non-compliance
follows. Compliance, in turn, is taken to mean that all measures are legally
valid. Then, the compensating measures are directly related to the AI Act.

The AI Act enumerates six conditions that must all be simultaneously satisfied
in order for a compliant high-risk AI system to process special categories of
personal data. Among these is a condition that a correction of a bias cannot be
done with synthetic or anonymized data, and a condition that all sensitive
personal data is deleted after a bias has been corrected.\footnote{~Article 10,
paragraphs (5)(a) and 5(e) in \textit{supra note} 9.} The former condition is
important because it implies that alternatives to personal data should be
considered already during a training phase of a high-risk AI system. If a
consideration indicates that alternatives are available, the GDPR does not apply
because neither anonymized data nor synthetic data is personal data. The latter
condition implies that sensitive personal data can only be used during training
after which it should be deleted. This mandate may be a problem because existing
studies have indicated that many AI models are continuously retrained after
their product launches.\footnote{~Oleksandr Kosenkov, Parisa Elahidoost, Tony
Gorschek, Jannik Fischbach, Daniel Mendez, Michael Unterkalmsteiner, Davide
Fucci, and Rahul Mohanani, `Systematic Mapping Study on Requirements Engineering
for Regulatory Compliance of Software Systems' $<$\url{https://arxiv.org/abs/2411.01940}$>$.} A lack of security (and privacy)
knowledge has also often been acknowledged as a problem among professional
software engineers.\footnote{~Hala Assal and Sonia Chiasson, ``Think Secure from
the Beginning': A Survey with Software Developers' \textit{Proceedings of the
  CHI Conference on Human Factors in Computing Systems (CHI)} (2019) Glasgow
$<$\url{https://doi.org/10.1145/3290605.3300519}$>$. \textit{Supra note}~20.} To
this end, it may be difficult to comply also with the security requirements of
the AI Act for processing sensitive personal data.\footnote{~Article 10(5)(c) in
\textit{supra note} 9.}  The same point applies regarding the
GDPR.\footnote{~Article 5(1)(f) and Article~32 in \textit{supra note} 11.}

In what follows, a potential bias and a subsequent discrimination are framed to
and contextualized with AI systems for recruitment of employees. Potential
biases in this particular context are well-recognized in existing
research.\footnote{~Alejandro Pe{\~{n}}a, Ignacio Serna, Aythami Morales and
Julian Fierrez, `FairCVtest Demo: Understanding Bias in Multimodal Learning with
a Testbed in Fair Automatic Recruitment' \textit{Proceedings of the 2020
  International Conference on Multimodal Interaction (ICMI)} (2020) virtual
event $<$\url{https://doi.org/10.1145/3382507.3421165}$>$. Kiran Kumar Reddy
Yanamala, `Dynamic Bias Mitigation for Multimodal AI in Recruitment Ensuring
Fairness and Equity in Hiring Practices' \textit{Journal of Artificial
  Intelligence and Machine Learning in Management} (2022) 6
$<$\url{https://journals.sagescience.org/index.php/jamm/article/view/169}$>$.}
The real or perceived biases, uncertainties, and associated fears have
frequently appeared also in media in recent years.\footnote{~Charlotte Lytton,
`AI Hiring Tools May Be Filtering Out the Best Job Applicants' (2024) BBC
$<$\url{https://www.bbc.com/worklife/article/20240214-ai-recruiting-hiring-software-bias-discrimination}$>$.}
Given this background, it is understandable why the AI Act explicitly remarks
that AI systems for recruitment, promotion, termination, and similar
work-related events are classified as being high-risk
scenarios.\footnote{~Recital 57 in \textit{supra note} 9.} In addition to
potential unjust consequences to people's livelihoods, the AI Act also notes
that in this context workers' rights and people's fundamental rights may be
threatened.\footnote{~Data protection is a fundamental right in the EU according
to Article~8 in Charter of Fundamental Rights of the European Union (2012) C
326/391.} The scenarios that are considered in the next Section~\ref{sec: scenarios}
are further framed to and contextualized with professional and ethical software
engineers who have detected biases during a software development of an AI system
for recruitment and try to address these by the means of using special
categories of personal data. Regarding virtue and professional ethics, it can
also be mentioned that ethical guidelines for software engineers and computing
professionals in general prohibit discrimination ``\textit{on the basis of age,
  color, disability, ethnicity, family status, gender identity, labor union
  membership, military status, nationality, race, religion or belief, sex,
  sexual orientation, or any other inappropriate factor}''.\footnote{~The
Association for Computing Machinery, `ACM Code of Ethics and Professional
Conduct' (2018) $<$\url{https://www.acm.org/code-of-ethics}$>$.} By implication,
they must comply with the noted six conditions in the AI Act. In addition, they
must pick one valid exception from the GDPR for processing sensitive personal
data of natural persons.\footnote{~Article 9(2) in \textit{supra note} 11.}

\section{Scenarios}\label{sec: scenarios}

\subsection{Invalid Consent}

Invalid consent is the easiest scenario. It has also been discussed
previously.\footnote{~\textit{Supra note} 3.} The essence is that a correction
of a bias by processing personal data, including sensitive personal data, is
justified in terms of the GDPR by obtaining consents from people whose personal
data is processed.\footnote{~Article 6(a) and Article~9(2)(a) in \textit{supra
  note} 11.} Although a consent is usually a valid way to process personal data,
the context of recruitment makes it invalid; the data protection regulators have
emphasized that under these settings a consent cannot be considered as freely
given due to imbalances of power between employees and
employers.\footnote{~European Data Protection Board, `Guidelines 05/2020 on
Consent Under Regulation 2016/679' (2020)
$<$\url{https://www.edpb.europa.eu/sites/default/files/files/file1/edpb_guidelines_202005_consent_en.pdf}$>$.}
Due to violating the GDPR's consent requirements, a non-compliance is thus
likely present.\footnote{~Article 4(11) in \textit{supra note} 11.}

\subsection{Age and Gender}\label{subsec: age and gender}

Discrimination based on age and gender, or both, is a common problem in many
European labor markets. Addressing a potential bias in this regard does not
delve deep into the AI Act and the GDPR because age and gender are not special
categories of personal data under these regulations. Yet, both appear in the
EU's anti-discrimination laws. Therefore, also the GDPR's notion of legal
obligations may well work as a legal basis.\footnote{~Article 6(c) in
\textit{supra note} 11.} Assuming that also the GDPR's other requirements are
satisfied, compliance can be reasonably assumed.

\subsection{Public Interests}\label{subsec: public interests}

Discrimination based on ethnicity too is a common problem in many European labor
markets. This scenario is different than the one in Section~\ref{subsec: age and
  gender} because ethnic origin is sensitive personal data according to the GDPR
and thus also the AI Act. However, the AI Act emphasizes that special categories
of personal data can be processed when there are substantial public interests
involved, which might seem a reasonable assumption in this
scenario.\footnote{~Recital 70 in \textit{supra note} 9.} The Act also points
directly to the corresponding exemption in the GDPR.\footnote{~Article 9(2)(g)
in \textit{supra note} 11.}  Assuming that all noted six conditions in the AI
Act and the GDPR's other obligations are satisfied, compliance can be again
reasonably assumed for the specific algorithmic bias correction purpose.

\subsection{Other Purposes}

A simple scenario would involve collecting sensitive personal data for
correcting a bias in recruitment but then using the data collected also for
other purposes. A typical scenario would involve using personal data later on
for marketing and advertisement purposes. This scenario would be non-compliant
in terms of both the AI Act and the GDPR. In terms of the former, it is
emphasized that special categories of personal data can be used in high-risk AI
systems only when it is ``\textit{strictly necessary for the purpose of ensuring
  bias detection and correction}''.\footnote{~Article 10(5) in \textit{supra
  note} 9.} The GDPR's purpose limitation conveys the same meaning; personal
data can only be collected for ``\textit{explicit and legitimate purposes and
  not further processed in a manner that is incompatible with those
  purposes}''.\footnote{~Article 5(1)(b) in \textit{supra note} 11.} In
addition, the AI Act's already noted requirement to delete sensitive personal
data after correcting a bias would be likely also violated.

\subsection{Data Sharing}

The age, gender, and ethnicity scenarios in the previous Sections~\ref{subsec: age
  and gender} and \ref{subsec: public interests} can be continued with a further
scenario involving sharing of sensitive personal data with third-parties. For
instance, a European software engineering group seeking to develop a unbiased AI
system for recruitment might share special categories of personal data with
overseas researchers specialized into bias-free AI. The GDPR allows personal
data sharing through its concepts of data controllers and processors. However,
non-compliance is present in this scenario because among the six conditions in the
AI Act is a prohibition of a transmission of sensitive personal data to other
parties.\footnote{~Article 10(5)(d) in \textit{supra note} 9.}

\subsection{Public Data}

A further scenario might involve collecting sensitive personal data from the public
Internet in order to correct a bias. For instance, a company developing an AI
recruitment system might collect people's curriculum vitae from personal
websites. As people often include photographs in their curriculum vitae, it
would then be possible to infer gender, age, and also sensitive categories such
as ethnicity from the photographs.\footnote{~\textit{Supra note} 24.} At first
glance, this scenario might satisfy the GDPR's exemption for processing sensitive
personal data in case a person has manifestly made it public.\footnote{~Article
9(2)(e) in \textit{supra note} 11.}  If the purpose would only be to correct
biases, it would seem that also the AI Act's requirements would be satisfied in
principle. However, the several administrative fines levied by the European data
protection authorities against the face-recognition company Clearview AI
indicate that the GDPR is still a problem in this scenario. Many of the fines were
justified by violations of the GDPR's lawfulness and transparency
principles.\footnote{~European Data Protection Board, ``Hellenic DPA Fines
Clearview AI 20 Million Euros' (2022)
$<$\url{https://www.edpb.europa.eu/news/national-news/2022/hellenic-dpa-fines-clearview-ai-20-million-euros_en}$>$.}
Thus, it seems fair to argue that non-compliance might be present in this
curriculum vitae scenario.

\subsection{Synthetic and Anonymized Data}

The AI Act's case of synthetic or anonymized data for bias correction is
interesting because it should be possible to \textit{ex~post} demonstrate a
necessity of collecting sensitive personal data instead. Thus, a non-compliance
scenario might involve someone else demonstrating later on that synthetic or
anonymized data would have been sufficient. For instance, academic AI
researchers might demonstrate in a peer reviewed publication that an AI
recruitment system implemented for a large governmental agency was non-compliant
because it involved unnecessarily collecting sensitive personal data, whereas,
in reality, the biases involved could have been corrected by census data and
related archival data collected by a country's national statistical agency. Time
will tell how the regulatory case of synthetic or anonymized data will play out
in practice.

\subsection{Proxy Variables}\label{subsec: proxy variables}

Finally, an important scenario involves discrimination even in case no sensitive
personal data was collected to begin with and the GDPR's principle of data
minimization was also otherwise followed.\footnote{~The principle is specified
in Article 5(1)(c) in \textit{supra note} 11.} This scenario arises because of
proxy variables, meaning that some other variable used in an AI system proxies a
discriminatory factor.\footnote{~Information Commissioner's Office, `What About
Fairness, Bias and Discrimination?' (2023)
$<$\url{https://ico.org.uk/for-organisations/uk-gdpr-guidance-and-resources/artificial-intelligence/guidance-on-ai-and-data-protection/how-do-we-ensure-fairness-in-ai/what-about-fairness-bias-and-discrimination/}$>$.}
For instance, a person's home street address might proxy ethnicity in case
ethnic minorities are segregated to some specific suburbans in a major European
city. It might also well be that even a person's name proxies his or her ethnic
origin. However, it is generally unclear how such scenarios should be
mitigated. One option would be to collect all discriminatory factors enumerated
in a national law of a member state, correlating these with all other variables
used in an AI recruitment system, and then either dropping or adjusting the
correlated variables, and deleting the sensitive personal data collected
afterwards. Though, this option would certainly conflict with the GDPR's data
minimization ideal. Therefore, an alternative would be so-called anonymized
recruitment whereby applicants are instructed to anonymize factors that are
known to be discriminatory. Although there are advantages and disadvantages,
such anonymized practices have been observed to reduce discrimination
particularly during the early stages of recruitment processes.\footnote{~Alain
Lacroux and Christelle Martin-Lacroux, `Anonymous R\'esum\'es: An Effective
Preselection Method?' \textit{International Journal of Selection and Assessment}
(2020) 28 $<$\url{https://doi.org/10.1111/ijsa.12275}$>$.} The practices are
also a good example more generally because they showcase that a bias correction
does not necessarily have to involve collection of sensitive personal data to
begin with.

\section{Conclusion}\label{sec: conclusion}

This short paper discussed algorithmic fairness in terms of discrimination, AI
systems for recruitment, and a few relevant EU laws. Of the scenarios discussed,
the public interest scenario in Section~\ref{subsec: public interests} is the
most important one. It demonstrates that the EU's new AI regulation has
addressed a concern that was present in terms of the GDPR. That said, it remains
to be seen how the regulators and legal systems interpret bias correction in
terms of public interests. Also the last scenario in Section~\ref{subsec: proxy
  variables} is worth emphasizing because it demonstrates that anonymization,
and not maximization of personal data collection, might be a solution in the
recruitment context. In general, the existing case law tends to emphasize a
necessity condition for personal data collection.\footnote{~Heinz Huber v
Bundesrepublik Deutschland (2008) C-524/06.} An adequate justification for
necessity is presumably particularly important when collecting and processing
sensitive categories of personal data. Therefore, it will be interesting to
observe how the necessity condition will be interpreted in relation to the
public interests for bias mitigation in AI systems. The concept of
proportionality is important in this regard. Finally, the scenario analysis also
demonstrates that there is no one-size-fits-all approach to compliance with the
EU laws. Therefore, further work is required to deduce whether some of the point
raised generalize to other contexts.

\end{document}